\documentclass[proof]{WileyASNA-v1}

\articletype{Article Type}%

\received{9 November 2021}
\accepted{10 November 2021}

\begin{document}

\title{Quasars physical properties study 
based on the medium-band photometric survey}

\author[1,2]{Kotov S.*}

\author[1,2]{Dodonov S.}

\author[1,2]{Grokhovskaya A.}

\authormark{Kotov S. \textsc{et al}}

\address[1]{\orgdiv{LSPEO}, \orgname{SAO RAS}, \orgaddress{\state{Karachay-Cherkessian Republic}, \country{Russia}}}

\address[2]{\orgname{IAA RAS}, \orgaddress{\state{St.Petersburg}, \country{Russia}}}

\corres{*Kotov Sergey,
\email{sss.kotov@mail.ru}}

\presentaddress{Russia, Karachay-Cherkessian Republic, Nizhnij Arkhyz, 369167}

\abstract{The study is about the quasars cosmological evolution. We used the medium-band photometric survey at the 1-m Schmidt telescope of the Byurakan Astrophysical Observatory. The HS47-22 field with an area of 2.38 square degrees has been selected. We have classified objects and composed a sample of quasars using broadband photometric data and morphological classification from the DECaLS survey \citep{DECaLS2019}, infrared photometry from the WISE survey \citep{WISE2018}, spectroscopy from the SDSS survey \citep{Paris2018}, ROSAT X-ray data \citep{Molthagen1997}, FIRST radio data \citep{Becker1995} and stellar parallax data from the GAIA survey \citep{Gaia2018}. We have compiled a sample of 682 quasars, determined their photometric redshifts from the medium-band photometric data, and calculated the absolute stellar magnitudes. The space density of quasars was calculated for different luminosity ranges at different redshifts using the $\lambda$-CDM model with $\Omega_m = 0.3$ and $\Omega_\lambda = 0.7$ \citep{Hogg1999}. In this paper, we present the comparison of our results with other works.}


\maketitle

\section{Introduction}\label{sec1}

There are significant dissimilarity in the evolution of the QSO (Quasi Stellar Objects, quasars) space density according to the optical and X-ray data \citep{Miyaji2000}. Moreover, large differences exist in the evolution of the quasars space density and their luminosity functions in the range $3 < z < 5$ according to the data of various optical surveys (SDSS -- Sloan Digital Sky Survey \citep{Paris2018}, COMBO-17 -- Classifying Objects by Medium-Band Observations \citep{Wolf2003}, ALHAMBRA -- Advanced Large, Homogenous Area Medium Band Redshift Astronomical Survey \citep{Matute2012}, COSMOS -- Cosmological Evolution Survey \citep{Masters2012}), which used varied methods of quasars sampling. These differences may be due to selection effects inherent in various quasar selection techniques. Since the mechanism of QSO emission is due to the matter accretion onto a supermassive black hole, a more detailed study of quasars in this redshift range will allow to clarify the evolution of supermassive black holes in the Universe.

\section{The problem of a representative quasars sample compiling}\label{sec2}

It is necessary to obtain a sample of quasars as homogeneous as possible and with maximum completeness in the survey field, studying the evolution of quasars in different luminosity and redshift ranges. It should be taken into account the following points to compile the sample:
\begin{itemize}
    \item The wide variety of quasars observed characteristics is explained by many factors, such as the rate of accretion onto a black hole, the structure of the AGN (Active Galactic Nucleus) central region, the orientation of the AGN to the observer, absorption and emission of the underlying galaxy, etc. All these factors have a considerable influence on the observed quasar luminosity.
    \item Searching for quasars with different spectral characteristics requires different selection techniques with different search criteria. To compile the most complete sample of quasars, selection according to different criteria should be carried out in parallel.
    \item Due to different spectral characteristics of quasars at different redshifts, selection effects of various kinds arise, the influence of which can be mistakenly interpreted as the quasars evolution themselves. Therefore, when compiling a sample for studying the evolution of quasars, it is required to minimize the dependence of selection effects on redshift.
\end{itemize}

\begin{figure}[t]
	\centerline{\includegraphics[width=78mm]{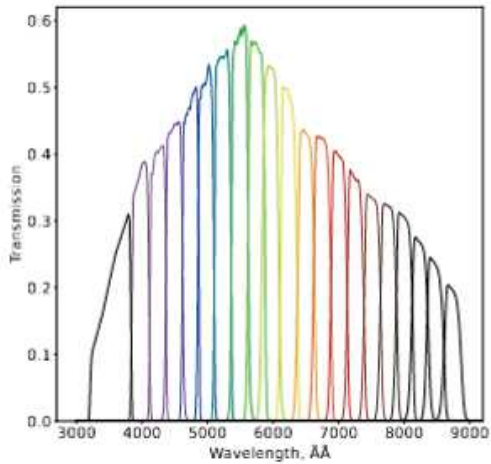}}
	\caption{The transmission curves of the medium-band filters, taking into account the detector sensitivity.\label{fig11}}
\end{figure}

\begin{figure*}[t]
\centerline{\includegraphics[width=472pt]{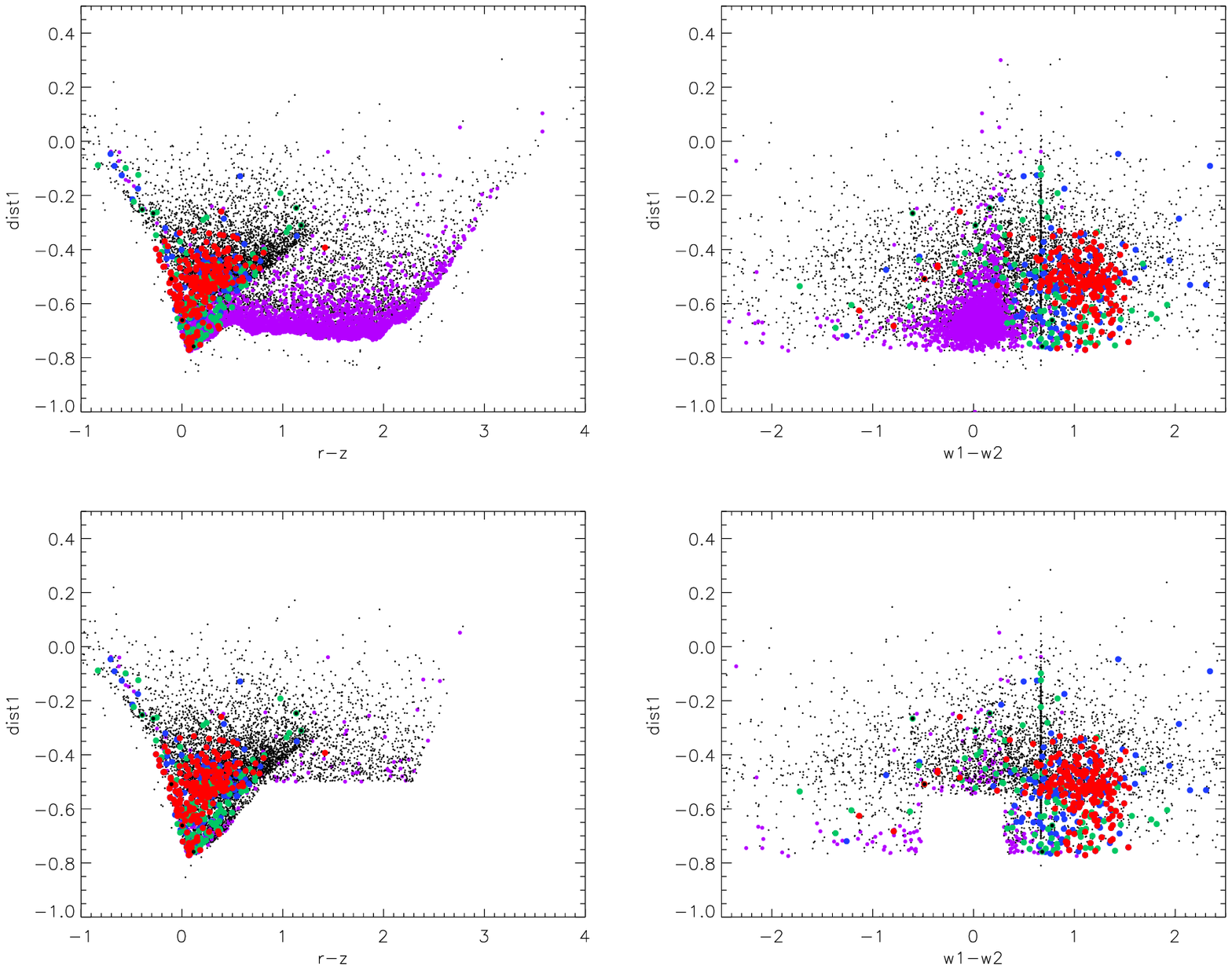}}
\caption{Combination of color space distance (dist1) with color criteria r-z (DECaLS \citep{DECaLS2019} and w1-w2 (WISE \citep{WISE2018}. The spectroscopically confirmed quasars are marked in red, and the quasar candidates selected manually from the medium-band photometric data are marked in blue and green. Objects with parallax according to Gaia data \citep{Gaia2018} are highlighted in purple. The upper figures show all the objects, the lower ones show the selected objects after the criteria have been applied.\label{fig1}}
\end{figure*}

\section{Data used to select quasars}\label{sec3}
We used the following data to search for quasars:
\begin{itemize}
    \item 1) Optical broadband photometry
    \item 2) Infrared photometry
    \item 3) Optical medium-band photometry
    \item 4) The presence of X-ray and radio emission
    \item 5) Morphological classification
    \item 6) Astrometric data: parallax and proper motions
    \item 7) Variability
    \item 8) Spectroscopic data
\end{itemize}

Optical broadband photometry does not require the use of large telescopes and makes it possible to obtain data on faint objects in the shortest telescopic time. Broadband photometry data make it possible to select quasars effectively at redshifts up to z = 2.2 by the ultraviolet excess criterion \citep{Richards2002}; however, at larger redshifts, quasars become practically indistinguishable from stars in broadband filters \citep{Kotov2016}. Also, it is impossible to determine the photometric redshifts of quasars with sufficient accuracy using these data.

Infrared color criteria allow to select most quasars at redshifts up to z = 3; at larger redshifts, the infrared colors of a significant part of quasars become indistinguishable from stars and galaxies \citep{IRcolors2013}. In addition, the currently available infrared data (the WISE survey, \cite{WISE2018} are not deep enough to select quasars fainter than the 22nd magnitude.

Optical medium-band photometry requires much more telescopic time in comparison with broadband photometry, but it allows to register broad emission lines in the QSO spectra \citep{Wolf2003}, which can be used for almost uniformly selection in the redshift range up to z = 5 or more. It is also possible to determine photometric redshifts of quasars with high accuracy using emission lines \citep{Chaves2017}.

The presence of X-ray and radio emission can be used as a criterion for the selection of quasars, but X-ray and radio emission sufficient for registration is observed only in a small part of quasars \citep{Silverman2008}, and optical identification of objects with follow-up spectroscopy is required to determine redshifts. Optical identification of X-ray objects is often ambiguous due to the low angular resolution of X-ray telescopes: several objects in optics can correspond to one X-ray source.

The morphological data can be used to separate quasars and distant galaxies.

Parallax and proper motion data can be used to separate stars from a sample. The deepest and most accurate data to date is the Gaia survey \citep{Gaia2016}. It presents data on parallaxes and proper motions for point sources, up to 20.5 magnitude \citep{Gaia2018}. This depth limit makes it possible to select only the brightest quasars at high redshifts.

The variability data can be used as a selection criterion, since the nature of the variability of quasars and variable stars is significantly different \citep{Palanque2016}. But obtaining such data implies many observations at different times. In addition, all processes at the redshift z for the observer occur $z + 1$ times slower \citep{Hogg1999}, and the variability of distant quasars become less detectable. This gives rise to redshift-dependent selection effects.

In most cases, spectroscopic data are exhaustive for objects classification and redshift determination; however, it require a long telescopic time and can be obtained only for pre-selected objects with precisely known coordinates in the field \citep{Paris2018}.

Since all the criteria individually have either selection effects depending on the redshift or affected from incompleteness, in order to obtain the most complete homogeneous sample, it is optimal to divide the selection into two stages. First, make the most complete, but heavily contaminated sample of QSO candidates, using various combinations of selection criteria in parallel. Then, remove contaminants from it using more sophisticated methods such as medium-band photometry and spectroscopy. Also, to determine the redshifts with an accuracy sufficient for studying evolution, either the data of medium-band photometry or spectroscopy data can be used.

Currently, there are many deep broadband photometric surveys covering most of the sky (SDSS \citep{Paris2018}, DECaLS \citep{DECaLS2019}, PanSTARRS \citep{Panstarrs2003}, their data is publicly available. Also in the public domain are data from the WISE infrared survey, data from the ROSAT X-ray telescope, the FIRST radio survey and the Gaia astrometric survey. Also, SDSS survey spectroscopy is available for many quasars.

To compile a complete sample of quasars, we chose the HS47-22 field, for which data from all the surveys described above are available. So it is possible to develop and apply a combined method for quasars selection. To maximize the quality of selection and redshift determination, the medium-band photometric survey in this field was made.

\section{Medium-band survey}\label{sec4}

\begin{figure}[t]
	\centerline{\includegraphics[width=78mm]{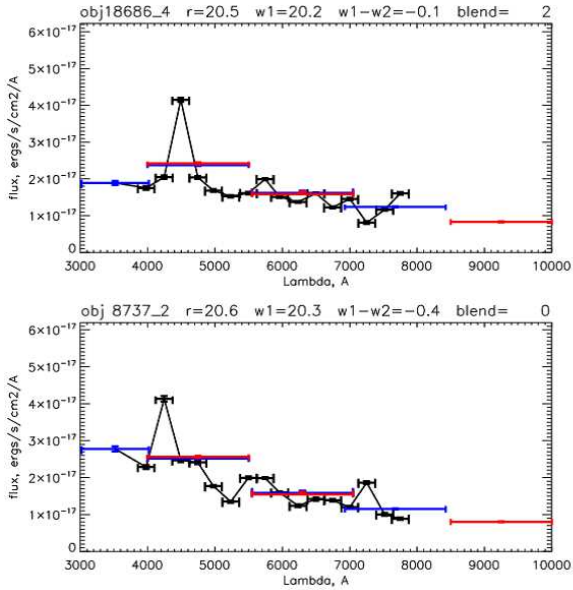}}
	\caption{Medium-band spectral energy distributions as a low-dispersion spectrum (black curve). Also the broadband photometry data from the 1-m Schmidt telescope (blue bars) and the DECaLS survey (red bars).\label{fig2}}
\end{figure}

A medium-band photometric survey in the HS47.5-22 field with an area of 2.386 sq. deg. was made on the 1-m Schmidt telescope of the Byurakan Astrophysical Observatory. The telescope was equipped with a 4k x 4k pixel CCD detector, which, at a scale of about 0.868" per pixel, gave a field of view of 1 square degree. The observations were carried out with 16 medium-band filters in range $4000 - 8000$ \AA$ $ and 4 broadband SDSS filters (u, g, r, i). The transmission curves of the medium-band filters, taking into account the detector sensitivity, are shown in the figure \ref{fig11}. The HS47 field was captured as a mosaic of 4 fields of 1 square degree with 10 arcmin overlaps, the final field size after all reductions was 2.38 square degrees. We achieved a signal-to-noise ratio of 5 for $AB = 23^m$ in all medium-band filters.

\section{Object selection technique}\label{sec5}

\begin{figure}[t]
	\centerline{\includegraphics[width=78mm]{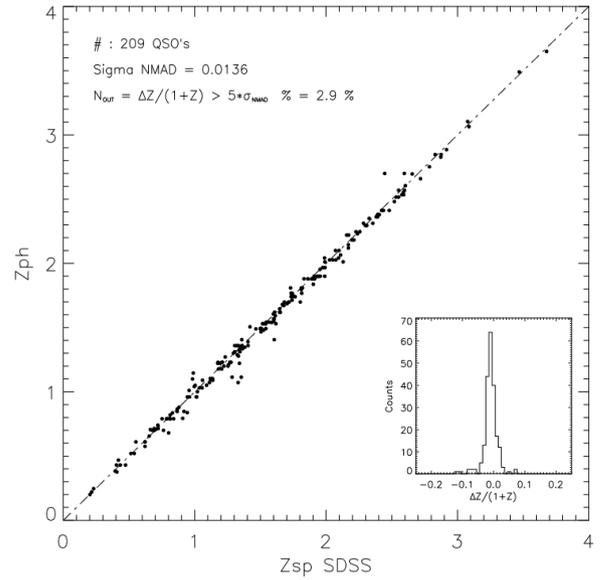}}
    \caption{The quality of the medium-band photometric redshifts determination, using data from 1-m Schmidt telescope. The vertical axis shows the photometric redshifts, the horizontal axis shows the redshifts according to the SDSS spectroscopy data (top figure). The accuracy of the determination was 0.02 with the amount of outliers about 9\% (bottom figure). \label{fig3}}
\end{figure}

Firstly, a coordinate-wise identification of objects with other surveys was made. We used WISE infrared photometry \citep{WISE2018}, DECaLS deep broadband photometry \citep{DECaLS2019}, morphology data from the DECaLS survey and parallax data from the GAIA survey \citep{Gaia2018}. We also identified the ROSAT X-ray sources \citep{Molthagen1997} and the FIRST radio sources \citep{Becker1995}.

We have studied the QSO morphology in the field according to DECaLS data. It showed that 97\% of known quasars are point sources, the remaining 3\% are classified as extended sources. At the same time, the DECaLS morphological classification proved to be excellent for the star-galaxy classification up to a magnitude of 22.2; at weaker values, distant compact galaxies are classified as point objects. Therefore, we used this as a selection criterion, which practically does not affect the completeness of the QSO sample, while allowing us to remove extended objects.

The next step was to develop an automatic selection algorithm. It is based on the idea that stars are numerous, and form a narrow sequence in the color space. On the other hand, quasars, especially at $z > 2.5$, are scattered in the color space much wider, and their quantity is by an order less. Therefore, we can consider for each object the distance to the nearest neighbors in the color space. For stars, this density will be significantly higher. This method proved to be excellent for objects brighter than $R = 20.5^m$ in the color space of medium-band filters. However, due to the noise, it cannot effectively separate stars at weaker values. In the space of broadband filters, the selection efficiency by this method turned out to be much lower for bright objects, but made it possible to significantly refine the sample of faint stars to $R = 22.5^m$. By combining the distance to nearest neighbors with DECaLS broadband colors and WISE infrared colors, we have obtained reliable criteria for separating stars and quasars (figure \ref{fig1}). Furthermore, according to Gaia DR2 data, objects with proper motions greater than $3\sigma$ level were excluded from the sample, because quasars are distant extragalactic sources and their proper motions can not be registered with GAIA DR2 precision. After that, the resulting sample of quasar candidates was supplemented with objects identified with X-ray and radio sources.

At the next stage, the medium-band spectral energy distributions of the sample objects were constructed, considered as low-dispersion spectra (figure \ref{fig2}). Each of them was looked through manually. All objects classified as quasars were selected in the final sample.

At the last stage, the photometric redshifts were determined using the ZEBRA package \citep{Zebra2006}, the correctness of the determination was checked manually. We have made spectroscopy of individual objects on the BTA-6 telescope (Large Altazimuth Telescope of SAO RAS), checking quality of classification and redshift determination. Figure \ref{fig3} shows the ratio of photometric and spectroscopic redshifts of objects. The determination accuracy was 0.02 with the amount of outliers about 9\%.

\begin{figure}[t]
	\centerline{\includegraphics[width=78mm]{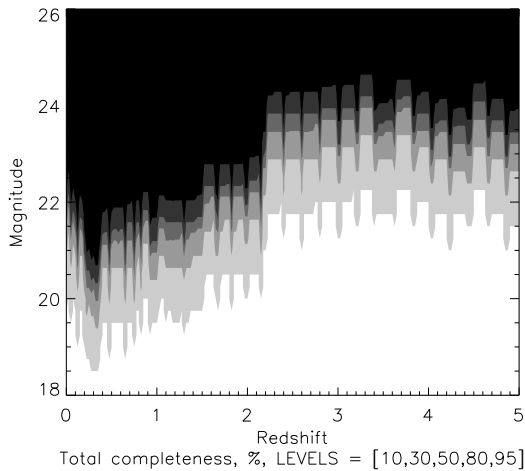}}
    \caption{Modeling the completeness of the quasars selection from medium-band photometric data. The color gradation from white to black indicates completeness levels of 95, 80, 50, 30 and 10 percent, respectively. \label{fig4}}
\end{figure}

\subsection{Quasar sample completeness estimation}

We have modeled the completeness of the QSO selection using the detection limit of emission lines in medium-band filters. The distribution of each emission line over the equivalent widths \citep{Chilingarian2003} and the broadening of the lines due to the redshift was taken into account. The line was considered registered if it was detected at the $3 \sigma$ level over the continuum. The calculation used 6 emission lines: $L_{\alpha}, CIV, CIII, MgII, H_{\beta}, H_{\alpha}$. It can be seen that at high redshifts ($z > 2.5$) the probability of detecting lines is estimated at 80\% and higher for objects brighter than $AB = 22.5^m$ (figure \ref{fig4}).

\section{QSO evolution study}\label{sec6}

\begin{figure}[t]
	\centerline{\includegraphics[width=78mm]{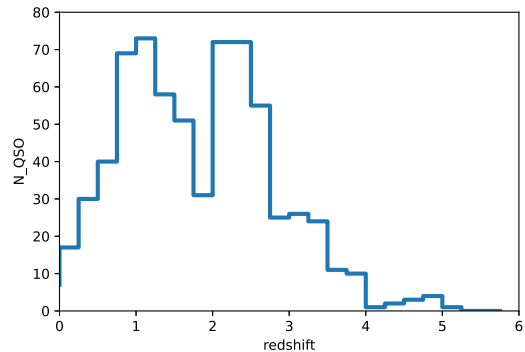}}
    \caption{The redshift distribution histogram for our sample. \label{fig0}}
\end{figure}

\begin{figure*}[t]
\centerline{\includegraphics[width=390pt]{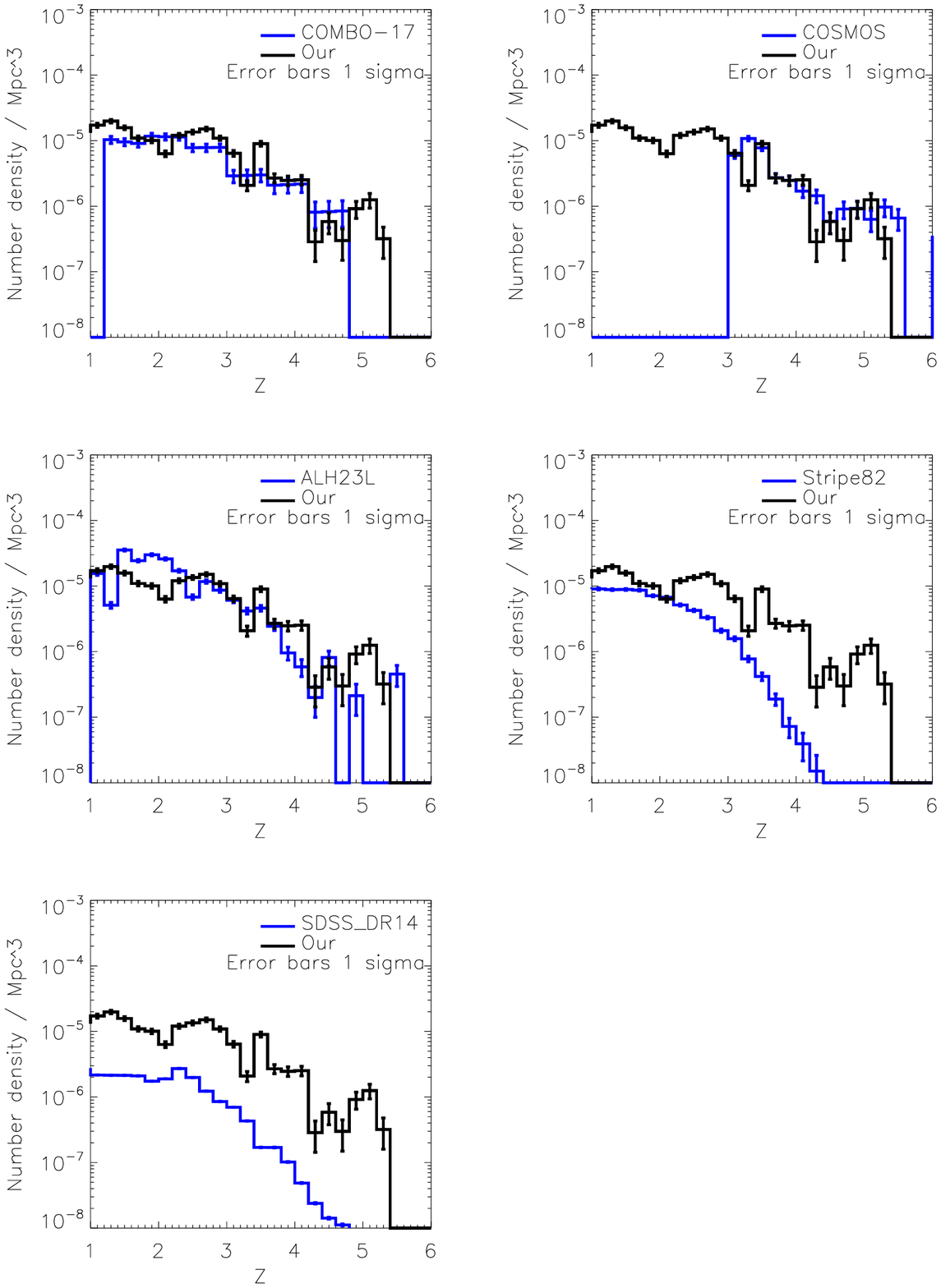}}
\caption{Comparison of the space density of quasars according to our data and data from other surveys (COMBO-17 \citep{Wolf2003}, COSMOS \citep{Masters2012}, ALHAMBRA \citep{Chaves2017}, Stripe-82 \citep{Palanque2016}, SDSS-DR14 \citep{Paris2018}.\label{fig5}}
\end{figure*}

Photometric redshifts were used to calculate the QSO space density and construct the luminosity function. The redshift distribution histogram is shown in figure \ref{fig0}. The commoving volume curvature was calculated within the $\lambda$-CDM model ($\Omega_m = 0.3, \Omega_\lambda = 0.7$) \citep{Hogg1999}, and a histogram of the QSO space density to redshift was constructed. The results were compared with those of other surveys. 

The absolute AB-magnitudes in the rest frame were calculated using redshift and space curvature within the $\lambda$-CDM model($\Omega_m = 0.3, \Omega_\lambda = 0.7$ \citep{Hogg1999}). We have compared the QSO space density calculated from our data and from the data of other surveys (figure \ref{fig5}). 

Our survey gives us 682 quasars in the 2.38 square degree survey field. We have used data from the following surveys for comparison:
\begin{itemize}
    \item SDSS-DR14 (selection based on broadband photometry data and infrared colors. 9376 sq. deg, 899098 quasars) \citep{Paris2018};
    \item Stripe-82 (selection based on variability and broadband photometry data) \citep{Palanque2016};
    \item ALHAMBRA (medium-band and broadband photometry. 2.79 sq. deg, 1079 quasars) \citep{Chaves2017};
    \item COSMOS (medium-band and broadband photometry, morphology. 2 sq. deg, 155 quasars on redshift $z > 3$) \citep{Masters2012};
    \item COMBO-17 (medium-band photometry. 0.78 sq. deg, 192 quasars) \citep{Wolf2003}). 
\end{itemize}

\begin{figure}[t]
	\centerline{\includegraphics[width=78mm]{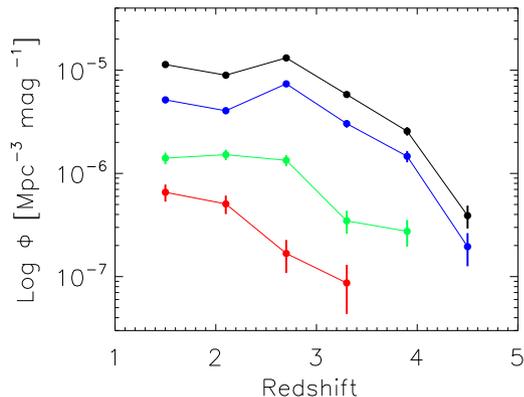}}
    \caption{Dependence of the QSO space density on the redshift, plotted for different luminosity ranges. Luminosity was calculated for the 145 nm wavelength in the restframe. Black is for quasars in the field brighter than $-24^m$; blue is for quasars brighter than $-25^m$; green is for quasars brighter than $-26^m$; red is for quasars brighter than $-27^m$.   \label{fig6}}
\end{figure}

We can see the difference in QSO space density according to different surveys. For SDSS-DR14, Stripe-82, ALHAMBRA, and COMBO-17 the space density begins to decrease in the redshift range $2 < z < 2.5$, while by our data the decrease begins only at $z = 3$. According to the COSMOS survey, in the range $3 < z < 5$, the distribution of the QSO space density coincides with our data.

The graph of the QSO space density binned by luminosity ranges (figure \ref{fig6}) clearly demonstrates that for objects brighter than $-24^m$, $-25^m$ and $-26^m$ the decrease in space density begins at $z > 3$. At the same time, for the brightest objects (brighter than -27m), the space density begins to decrease significantly earlier, at $z = 2.5$, This can be explained both by the evolution of the QSO luminosity and by random fluctuations, since the size of the bright quasars sample is quite small (several tens of objects).

\section{Conclusions}\label{sec7}

According to our data, we obtained the dependence of the QSO space density on redshift, which is in good agreement with the data of the COSMOS survey in the redshift range $3 < z < 5$. According to our data, the decrease in the QSO space density begins at z = 3.0, which is in good agreement with the X-ray data \citep{Miyaji2000}. At the same time, in the COMBO-17 survey, the decline begins already at $z = 2.6$, in the ALHAMBRA survey – at $z = 2.4$, in the SDSS-DR14 survey - at $z = 2.5$, and in the Stripe-82 survey - at $z = 2.0$. We associate these differences with the fact that with an increase in the depth and an improvement in the quality of QSO selection, the influence of selection effects decreases, and the completeness of the sample at large redshifts increases. Thus, it can be assumed that a further increase in the depth and a decrease of selection effects influence will shift the maximum of the QSO space density towards larger redshifts.

\newpage


\section*{Acknowledgments}

We thank our colleagues of the SAO RAS and BAO NAN who helped us in 1-m Schmidt telescope modernisation and made this observations possible. We thank teams of SDSS, DECaLS, WISE, FIRST, ROSAT and GAIA surveys for open-access data which we used in our research. Observations with the SAO RAS telescopes are supported by the Ministry of Science and Higher Education of the Russian Federation (including agreement No05.619.21.0016, project ID RFMEFI61919X0016). The renovation of telescope equipment is currently provided within the national project "Science".

\subsection*{Author contributions}

Participation in the medium-band photometric survey on the 1-m Schmidt telescope of the BAO NAN;
Processing images obtained with the 1-m Schmidt telescope, with co-authors;
Development of methods for automatic QSO selection;
Manual selection of quasars by spectral energy distributions, with co-authors;
Modeling the QSO selection completeness; 
Calculation of the QSO space density, carrying out k-correction, taking into account the intergalactic absorption, constructing the QSO luminosity function.

\subsection*{Financial disclosure}

This work was partially supported by the \fundingAgency{Russian Science Foundation} grant \fundingNumber{21-12-00210}

\subsection*{Conflict of interest}

The authors declare no potential conflict of interests.

\section*{Supporting information}

The following supporting information is available as part of the online article:

\bibliography{Sergey_Kotov}
\section*{Author Biography}

\begin{biography}{}{\textbf{Sergey Kotov} Junior researcher of LSPEO, SAO RAS, and Junior researcher of LEA, IAA RAS. Experienced in optical surveys and physical properties of quasars.}
\end{biography}
\begin{biography}{}{\textbf{Sergey Dodonov} Director of the Laboratory for Spectroscopy and Photometry of the Extragalactic Objects, SAO RAS, and senior researcher of LEA, IAA RAS. Specialist in the optical surveys, spectroscopy and photometry, instrumentation for the telescopes. Russia State Prize Laureate.}
\end{biography}
\begin{biography}{}{\textbf{Alexandra Grokhovskaya} Junior researcher of LSPEO, SAO RAS, and Junior researcher of LEA, IAA RAS. Experienced in ML methods in astronomy.}
\end{biography}

\end{document}